\documentclass[%
 reprint,
 amsmath,amssymb,
 aps,
prb,
]{revtex4-2}

\usepackage{graphicx}% Include figure files
\usepackage{dcolumn}% Align table columns on decimal point
\usepackage{bm}% bold math
\usepackage{here}
\usepackage{layout}
\usepackage{mathtools}
\usepackage{afterpage}
\begin{document}

\preprint{APS/123-QED}

\title{Berezinskii - Kosterlitz - Thouless transition 
in rhenium nitride films}% Force line breaks with \\

\author{Kosuke Takiguchi}
 \email{kosuke.takiguchi@ntt.com}
\author{Yoshiharu Krockenberger}
\author{Yoshitaka Taniyasu}
\author{Hideki Yamamoto}
\affiliation{%
NTT Basic Research Laboratories, NTT Corporation, 3-1 Morinosato-Wakamiya, Atsugi, Kanagawa 243-0198, Japan
}%

\date{\today}% It is always \today, today,
             %  but any date may be explicitly specified

\begin{abstract}
  The quest to manipulate and understand superconductivity demands exploring diverse materials and unconventional behaviors. Here, we investigate the BKT transition in synthesized ReN$_x$ thin films, demonstrating their emergence as a compelling platform for studying this pivotal phenomenon. By systematically varying synthesis parameters, we achieve ReN$_x$ films exhibiting a BKT transition comparable or even surpassing the archetypal NbN$_x$ system. Detailed current-voltage measurements unlock the intrinsic parameters of the BKT transition, revealing the critical role of suppressed superconducting volume in pushing ReN$_x$ towards the two-dimensional limit. Utilizing this two-dimensional electron system, we employ Beasley-Mooij-Orlando (BMO) theory to extract the vortex unbinding transition temperature and superelectron density at the critical point. Further confirmation of the BKT transition is obtained through temperature-dependent resistivity, current-voltage, and magnetoresistance measurements. Our findings suggest that native disorder and inhomogeneity within ReN$_x$ thin films act to suppress long-range coherence, ultimately driving the system towards the BKT regime. This work establishes ReN$_x$ as a promising material for exploring BKT physics and paves the way for tailoring its properties for potential applications in superconducting devices.
\end{abstract}

%\keywords{Suggested keywords}%Use showkeys class option if keyword
                              %display desired
\maketitle

%\tableofcontents

    % For two-column output uncomment the next line and choose [10pt] rather than [12pt] in the \documentclass declaration
    %\ioptwocol
    %

    \newpage

    \section{Introduction}

    The pursuit of understanding and manipulating superconductivity continues to drive innovation across physics and material science. In this realm, superconducting thin films offer fertile ground for exploring diverse phenomena due to their controllable geometry and sensitivity to external factors. Among these, the Berezinskii-Kosterlitz-Thouless (BKT) transition \cite{Berezinsky1972, Kosterlitz1973, Kosterlitz1974} holds a unique position, unveiling the interplay between order and disorder within the vortex condensate. \par
    In essence, the BKT transition marks a crucial shift from a state with quasi-long-range order in the vortex world to a fully disordered regime as temperature increases. This transition, originally conceptualized for two-dimensional (2D) systems, finds profound relevance in thin films where finite thickness creates an effective 2D superconducting layer. Studying the BKT transition in these systems unlocks valuable insights into fundamental aspects of vortex dynamics, phase fluctuations, and the interplay between dimensionality and superconductivity.\par
    The importance of the BKT transition stems from its multifaceted influence on superconducting properties. Its influence extends beyond a simple phase change, impacting a variety of observable characteristics. Regarding electronic transport, the characteristic discontinuities in resistivity near the BKT temperature $T_\mathrm{BKT}$ unveil the breakdown of phase coherence and the emergence of dissipative processes. The magnetization behavior of the BKT transition leaves its footprint in the hysteresis behavior of magnetization, reflecting the collective dynamics of unbound vortices. Furthermore, the BKT transition plays a pivotal role in understanding the interplay between superconductivity and other phase transitions, particularly the superconductor-to-insulator transition. By tuning film thickness, disorder, and external fields, researchers can manipulate the BKT transition and explore this intriguing interplay, paving the way for novel device functionalities \cite{Mondal2011,Mondal2012,Yong2013,Venditti2019,Weitzel2023,Bartolf2010,Baturina2012,Nguyen2020}. \par 
    In light of these multifaceted implications, a meticulous investigation of the BKT transition in superconducting thin films remains an active and impactful area of research. This manuscript delves into the intricacies of this fascinating phenomenon, shedding light on its mechanisms, consequences, and potential applications. Through detailed analysis of experimental data and theoretical frameworks, we aim to deepen our understanding and pave the way for further exploration of this remarkable transition in the fascinating world of superconducting thin films. \par

    For this purpose, the choice of the experimental platform is crucial. Homogeneously disordered superconducting thin film materials are widely studied for the BKT physics though a clear distinction between three-dimensional superconductivity and 2D BKT transitions appears to be challenging. In fact, truly long-range homogeneously disordered thin film materials without superconducting domains are a major issue for experimental characterization of the BKT transition. The NbN$_x$ system is a favorite to investigate the BKT phase transition. Yet, several shortcomings associated to it need to be addressed. One of the major issues is that NbN$_x$ can be synthesized with long-range crystalline coherence using standard synthesis techniques such as radio frequency (RF) sputter \cite{Chockalingam2008}. As such long-range coherence of crystalline phases is not desirable for the investigation of the BKT transition, the films have to extremely thin ($<$ 3\,nm) \cite{Mondal2011,Mondal2012,Yong2013,Venditti2019, Weitzel2023} or subject to Ar ion bombardment \cite{Bartolf2010}. After subjecting NbN$_x$ to Ar ion plasma, NbN$_x$ may partially dissociate, leaving Nb instead. However, Nb is a superconductor by itself with a superconducting transition temperature nearby that of NbN$_x$. Similar experimental requirements have been reported also for TiN films \cite{Baturina2012}. \par
    Here, we introduce the ReN$_x$ system as an alternative materials system for the investigation of the BKT transition. Compared with the NbN$_x$ and the TiN system, the ReN$_x$ system is not in favor of long-range crystallographic coherence and high-pressure-high-temperature synthesis techniques are required. Such a thermodynamic behavior is beneficial for the investigation of the BKT transition as the as-grown material is sufficient to mimic a large-scale Josephson junction system \cite{Wees1987}. In the ReN$_x$ system, superconducting (ReN\cite{Haq1983,Fuchigami2009} and ReN$_2$\cite{Onodera2019}) and non-superconducting phases (Re$_3$N \cite{Friedrich2010}, Re$_7$N$_3$\cite{Dubrovinsky2022}, Re$_2$N \cite{Friedrich2010}, Re$_3$N$_2$\cite{Zhao2014}, ReN$_3$\cite{Zhao2014}, ReN$_4$\cite{Zhao2014} and ReN$_{10}$\cite{Zhao2014}) are known. The variety of ReN$_x$ phases reported originates from the covalent bonding character of Re and N \cite{Zhao2014}. In contrast, for NbN - a 4$d$ nitride system - the ionic character increases. Notably, ionic character for TiN - a 3$d$ system - appears to be highest. As the ReN$_x$ system easily enables coexisting superconducting and non-superconducting phases, introduction of the necessary disorder appears to be easier than in the NbN$_x$ system. \par

    In this paper, we report the BKT transition in ReN$_x$ films. Magnetization measurements using a superconducting quantum interference device (SQUID) on ReN$_x$ films show a transition from fully diamagnetic thin films towards ReN$_x$ films with a paramagnetic response. Using the theory of Beasley, Mooij, and Orlando (BMO)\cite{Beasley1979}, we confirm the BKT transition by temperature dependence of resistivity, non-linear current-voltage characteristics, and magnetoresistance data even in the sample with the smallest superconducting volume. Our results on the BKT behavior of the ReN$_x$ system show that it can be considered as an alternative model system for the investigation of the BKT transition.

    \section{Experimental details}
    ReN${_x}$ thin films were synthesized on KTaO$_3$(001) substrates by molecular beam epitaxy. We synthesized five samples: (I) 7.8-nm-thick ReN$_x$ grown at $T_s =$ 500\,$^\circ$C, (II) 7.8-nm-thick ReN$_x$ grown at 600\,$^\circ$C, (III) 7.8-nm-thick ReN$_x$ grown at 700\,$^\circ$C, (IV) 19.5-nm-thick ReN$_x$ with 1.5 mol \% praseodymium (Pr) grown at $T_s =$ 700\,$^\circ$C, and (V) 39-nm-thick ReN$_x$ with 1.5 mol \% Pr grown at $T_s =$ 700\,$^\circ$C. $T_s$ represents the growth temperature. A custom-designed nitrogen RF source is used for the nitrification of Re with a typical N$_2$ flow rate $f_\mathrm{N_2}$ of 2.0 sccm. The RF radical power $P_\mathrm{RF}=$ 400\,W in samples I-III, and 350\,W in samples IV and V. Detailed information of the sample growth and structure determination are given in \cite{Sup}. Re$_2$N is a non-superconducting phase in the ReN$_x$ system and the synthesis conditions used determine the ratio between superconducting ReN and non-superconducting Re$_2$N \cite{Friedrich2010}. Higher synthesis temperatures favor the formation of ReN. If very high synthesis temperatures are required, the introduction of a catalyst appears to be beneficial. It is desired that the catalysts electronic behavior is insulating and we therefore deposited praseodymium \cite{Abu-Zied2015, Liu2022}. Under flowing atomic nitrogen, rare earth elements readily form rare earth nitrides, which are electronically insulating \cite{Sclar1964}. Addition of Pr enables an additional control parameter of the superconducting volume fraction associated with ReN. \par 
    Samples I-IV were used for transport and magnetization measurements. Additional data taken of sample III and V are shown in \cite{Sup}. Resistivity ($\mathrm{\rho}$) measurements were performed using a standard four-probe method with 40\,nm silver electrodes. Temperature dependence of $\mathrm{\rho}$, electrical current-voltage ($\mathrm{V}(I)$) characteristics, and magnetoresistivity were measured in a Quantum Design Dynacool PPMS under magnetic fields applied perpendicular to the film surface. Electronic transport measurements between 100 mK and 1.8 K are carried out using a $^3$He$-^4$He dilution refrigerator option on Quantum design PPMS.

    \section{Results and discussion}
    \subsection{Temperature dependence of magnetic moment}
\begin{figure}[t]

      \includegraphics[width=\columnwidth]{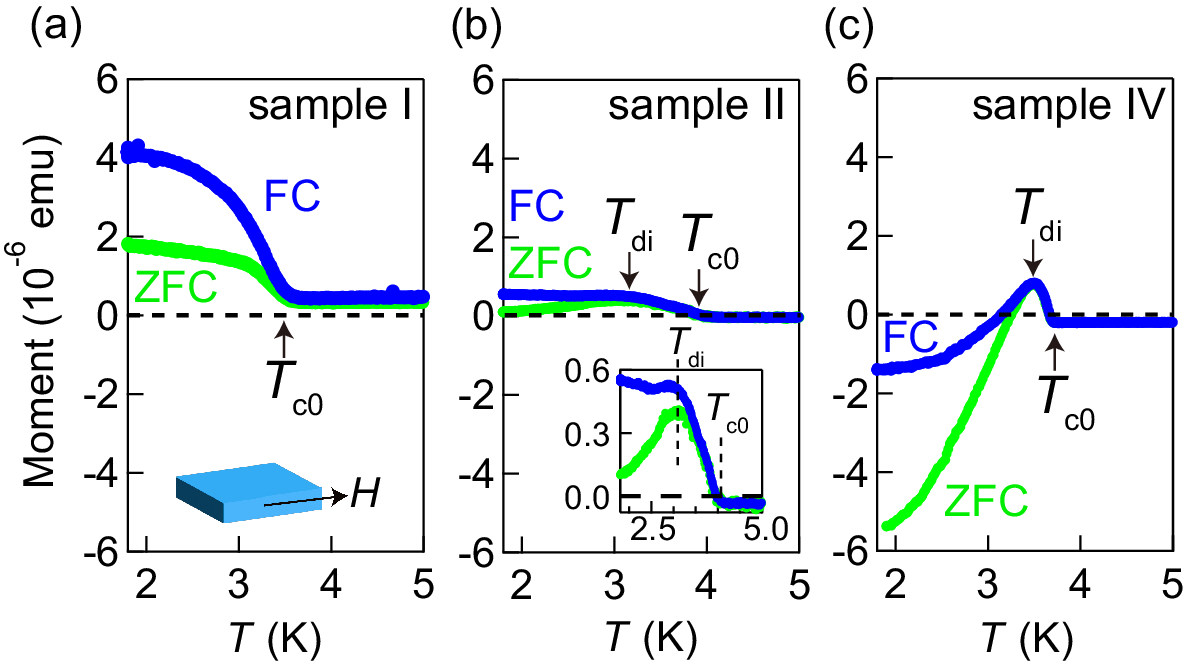}
      \caption{Temperature dependence of magnetic moment of samples (a) I (b) II, and (c) IV. The applied magnetic field of $H$ = 10 Oe is parallel to the in-plane direction of the film. For each sample shown here, field-cooled (FC) and zero field cooled (ZFC) measurement data are plotted. The inset of (b), a magnified view of the magnetization in the vicinity of $T_\mathrm{di}$ and $T_\mathrm{c0}$ is plotted. The dashed line indicates the zero moment. Note that the magnetization remains positive at all temperatures for samples I and II. On the other hand, shielding (ZFC) and Meissner (FC) diamagnetic signals are clearly observed in (c).}
      \label{fig:MT}

    \end{figure}

    Figures \ref{fig:MT}(a-c) show the temperature dependence of magnetic moment for samples I, II, and IV. An external magnetic field of $H=$ 10\,Oe was applied in the in-plane direction of the film. For sample I [Fig. \ref{fig:MT}(a)], one can see the transition to the positive magnetic signal below $T_{\mathrm{c0}}=$ 3.53\,K in both field cooling (FC) and zero field cooling (ZFC). $T_{\mathrm{c0}}$ represents the mean field critical temperature. Since the formation of Cooper pairs is hindered due to the inhomogeneity, superelectrons yields a paramagnetic signal in the temperature range of the BKT phase. In the previous reports, the positive magnetic response is typical in a Josephson-junction array or superconducting thin films in the two-dimensional limit \cite{Koblischka2023}. The positive response in FC is larger than that in ZFC, and this is a different feature of Meissner effect in superconductors. In sample II, this behavior is also observed at $T_{\mathrm{c0}}=$ 4.00\,K, and the moment below $T_{\mathrm{c0}}$ is suppressed rather than that of sample I. The diamagnetic signal becomes greater than the positive signal from the BKT state below $T_\mathrm{di}=$ 3.10\,K. $T_\mathrm{di}$ is temperature below which the magnetic moment decreases in ZFC. The inflection points in FC coincides with $T_\mathrm{di}$, implying that the transitions seen in FC and ZFC occur from the same origin. The coherency between superelectrons develops below $T_\mathrm{di}$. The situation, in which the diamagnetic signal overcomes the positive magnetic response by BKT state, is more pronounced in sample IV. Meissner effect is clearly seen in FC and ZFC, respectively. When 3.48\,K (= $T_{\mathrm{di}}$) $<T<$ 3.70\,K ($=T_{\mathrm{c0}}$), the positive magnetic response is observed as well as in samples I and II. On the other hand, below $T_{\mathrm{di}}$, Meissner effect dominates the temperature dependence of the moment due to the large superconducting volume. We confirmed this by doubling the thickness of ReN$_x$ (sample V, see \cite{Sup}), and x-ray diffraction data confirm the existence of ReN$_x$ with $x\simeq1$. In contrast, large crystalline volumes are not seen for other samples discussed here. \par

    \begin{figure}[ht]

      \includegraphics[width=\columnwidth]{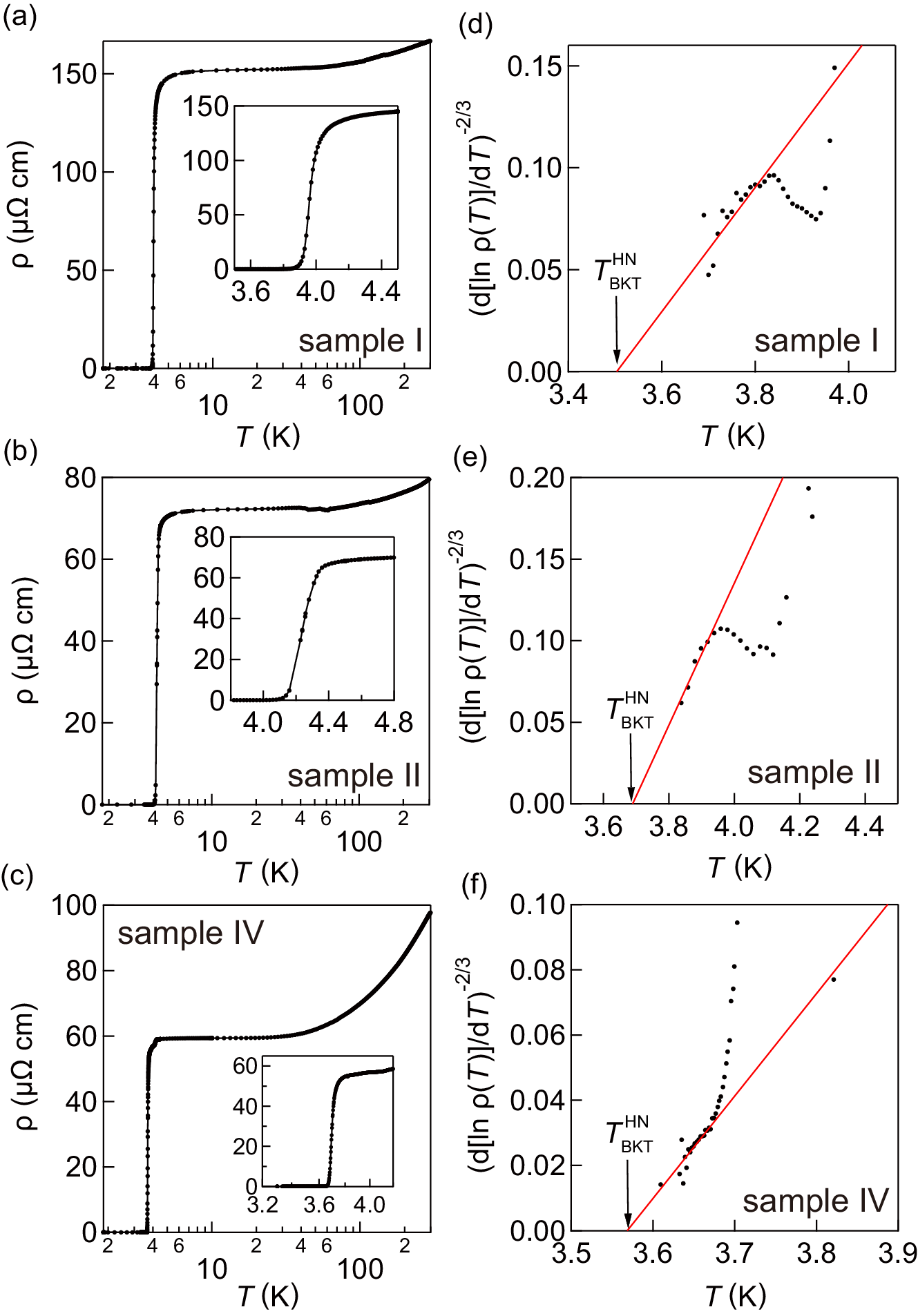}
      \caption{In Figs. (a), (b), and (c), the resistivity of samples I, II, and IV are plotted as a function of temperature (logarithmic scale). The insets in (a)-(c) enlarge on the resistivity drop to $\mathrm{\rho}=0$ around $T_\mathrm{di}$. Note that the temperature is plotted linearly. In (d)-(f), we plot $\left(\mathrm{d}\left[\ln \rho(T) \right]/\mathrm{d}T\right)^{-2/3}$ as a function of temperature for $\mathrm{\rho}$ in the vicinity of $T_\mathrm{di}$. The straight fitting curve intercepts $\left(\mathrm{d}\left[\ln \rho(T) \right]/\mathrm{d}T\right)^{-2/3}=0$ to reveal $T^\mathrm{HN}_{\rm{BKT}}$.}
      \label{fig:RT_HN}

    \end{figure}
    \begin{table*}[ht]
      \caption{Growth condition, electronic transport parameters and magnetization parameters determined for samples I, II, and IV. $\mathrm{\rho}_N$ is the resistivity obtained at 5.0\,K in each sample.}
      \label{table:Tc}
      \centering
      \begin{tabular}{l|c|c|c|c|c|c|c|c|c|c}
        \hline
        Sample  & $T_s$ ($^\circ$C)  &  Pr (\%) & $f_\mathrm{N_2}$ (sccm) & $P_\mathrm{RF}$ (W) & $T_{\mathrm{c0}}$ (K) & $\xi_0$ (nm) & $\rho_N \,(\mathrm{\mu \Omega\,cm})$&$\ell$ (nm) &$\mu_0H_{c2}(0)$ (T)&  $\xi_\mathrm{GL}(0) $ (nm)  \\
        \hline \hline
        I  & 500 & 0 & 2.0 & 400 & 3.54& 590 & 152 $\pm$ 9.83 &0.47 $\pm$ 0.03 & 3.55 $\pm$ 0.008& 9.63 $\pm$ 0.01  \\
        II& 600 & 0 &  2.0 & 400 & 4.00& 522& 72.7 $\pm$ 4.69&0.98 $\pm$ 0.06& 3.83 $\pm$ 0.014& 9.27 $\pm$ 0.02 \\
        IV& 700 &  1.5 & 2.0 & 350 & 3.70& 565& 39.4 $\pm$ 1.06 &1.80 $\pm$ 0.05 & 1.51 $\pm$ 0.004& 14.8 $\pm$ 0.02\\
        \hline
      \end{tabular}
    \end{table*}

    \begin{table*}[ht]
      \caption{BKT parameters determined for samples I, II, and IV.  $J_s$, $n_s^{\mathrm{2D}}$, and $\lambda_{\perp}$ were estimated by using $T^\mathrm{HN}_\mathrm{BKT}$ according to $J_s=2T^\mathrm{HN}_\mathrm{BKT}/\pi$, Eq. (\ref{eq:BMO_n}), and Eq. (\ref{eq:BMO_l}), respectively. }
      \label{table:BKT}
      \centering
      \begin{tabular}{l|c|c|c|c|c}
        \hline
        Sample  &$T^{\mathrm{HN}}_{\mathrm{BKT}}$ (K)&$T^{\alpha}_{\mathrm{BKT}}$ (K) &$J_s$ (K)&$n_s^{\mathrm{2D}}$ ($10^{15}$ m$^{-2}$)& $\lambda_{\perp}$ (mm)\\
        \hline \hline
        I  &3.50 $\pm$ 0.48&3.46&2.23 $\pm$ 0.31&5.04 $\pm$ 0.69&2.80 $\pm$ 0.38\\
        II &3.69 $\pm$ 0.61&3.52&2.35 $\pm$ 0.39&5.31 $\pm$ 0.88&2.66 $\pm$ 0.44\\
        IV &3.57 $\pm$ 0.66&3.63&2.27 $\pm$ 0.42&5.14 $\pm$ 0.95&2.75 $\pm$ 0.51\\
        \hline
      \end{tabular}
    \end{table*}

    \subsection{Temperature dependence of resistivity}

    Next, we show temperature dependence of resistivity ($\mathrm{\rho}$) [Fig. \ref{fig:RT_HN}(a-c)] of samples I, II, and IV, respectively. The normal state resistivity at 5.0 K $\mathrm{\rho}_N$ are shown in Table \ref{table:Tc}. From $\mathrm{\rho}_N$, we can discuss the inhomogeneity of our ReN$_x$ by employing the NbN data in the previous report since NbN has the same crystal structure of ReN. Using a Fermi velocity $v_{F, \mathrm{NbN}}=1.52 \times 10^6$\,m/s and carrier concentration $n_\mathrm{NbN}=7.60\times10^{28}$\,m$^{-3}$ equal to what have been reported for NbN \cite{Chockalingam2008}, the BCS coherence length $\xi_0(=0.18 \hbar v_{F, \mathrm{NbN}}/k_BT_\mathrm{c0})$ and the mean free path $\ell(=mv_{F, \mathrm{NbN}}/e^2\rho_Nn_\mathrm{NbN})$ are also calculated (Table \ref{table:Tc}), where $k_B$ is Boltzmann constant, $m$ is electron mass, and $e$ is elementary charge. $\ell$ becomes longer in the order of samples I, II, and IV as the crystallized volume increases. All $\ell$ in samples I, II, and IV are shorter than 0.3\% of $\xi_0$. Hence, the dirty limit condition is fulfilled ($\ell\ll\xi_0$), and the disordered state for the BKT phenomenon is realized. Near $T_\mathrm{BKT}$, spontaneous creation and annihilation of vortex-antivortex pairs terminates long-range coherency and zero resistance emerges at temperatures where fluctuations are sufficiently suppressed. This transition corresponds to the BKT transition. According to the theory of Halperin-Nelson (HN) \cite{Halperin1979} on the BKT transition, the $\mathrm{\rho}(T)$ behavior in this region follows
     \begin{equation}
      \label{eq:HN}
      T-T_{\rm{BKT}} \propto \left(\frac{\mathrm{d}\left[\ln \rho(T) \right]}{\mathrm{d}T} \right)^{-\frac{2}{3}}.
     \end{equation}
    As shown in Fig. \ref{fig:RT_HN}(d-f), we fitted the $\mathrm{\rho}(T)$ data using Eq. (\ref{eq:HN}). The fitting intervals of samples I, II, and IV are 3.70\,K $<T<$ 3.84\,K, 3.84\,K $<T<$ 3.94\,K, and 3.61\,K $<T<$ 3.67\,K, respectively. The estimated transition temperature ($T^\mathrm{HN}_{\rm{BKT}}$) and superfluid stiffness $J_s=2T^\mathrm{HN}_\mathrm{BKT}/\pi$ are shown in Table \ref{table:BKT}. According to the BMO theory \cite{Beasley1979}, the relation between $T_\mathrm{BKT}$ and the areal superelectron density $n_s^{\mathrm{2D}}$ is given by
    \begin{equation}
      \label{eq:BMO_n}
      k_BT_\mathrm{BKT}=\frac{1}{2}\frac{\pi\hbar^2n_s^{\mathrm{2D}}}{m^*},
    \end{equation}
    where $m^*$ is the particle mass which is $2m$ for superconductors. Eq. (\ref{eq:BMO_n}) gives $n_s^{\mathrm{2D}}$, as shown in Table \ref{table:BKT}. In terms of the penetration depth $\lambda$, $n_s^{\mathrm{2D}}$ satisfies
    \begin{equation}
      \label{eq:BMO_l}
      n_s^{\mathrm{2D}} = \frac{1}{2}\frac{mc^2}{4\pi e^2}\frac{d}{\lambda^2}=\frac{1}{2}\frac{mc^2}{4\pi e^2}\frac{1}{\lambda_\perp},
    \end{equation}
    where $d$ is the thickness. Beyond the characteristic distance $\lambda_{\perp}=\lambda^2/d$, proposed by Pearl \cite{Pearl1964}, the logarithmic interaction by the vortices pairs falls off as $1/r$, where $r$ is the distance between the vortices pairs. As shown in Table \ref{table:BKT}, for all the samples, we find $\lambda_{\perp}$ has same order magnitude of the sample dimensions ($\sim$2.5\,mm $\times$ 5.0\,mm). We therefore conclude that the vortex coupling strength is sufficient to stabilize a BKT state in two-dimensional ReN$_x$ films. \par

    \subsection{$\mathrm{V}(I)$ characteristics}
    To ascertain the absence of incidental long-rang order near the BKT transition, data from $\mathrm{V}(I)$ measurements are used. The driving mechanisms at the BKT transition require that large currents decouple vortex-antivortex pairs \cite{Nelson1977}. Then, extra electrical resistance $R$ is generated due to the free vortex. Let $n_v$ be the density of state of free vortex. The extra resistance follows $R = V/I \propto n_v$. It is known that $n_v$ scales with a power law of $I$. The exponent of $I$ is proportional to superfluid stiffness $J_s(T)$ and $T^{-1}$, where $m$ is electron mass. Thus, $\mathrm{V}(I)$ characteristics around BKT transition is described as
    \begin{equation}
      \label{eq:IV}
      V \propto I^\alpha , \alpha = 1+ \frac{\pi J_s(T)}{T}.
    \end{equation}
    When the BKT transition occurs ($T=T^\alpha_\mathrm{BKT}$), $J_s(T^\alpha_\mathrm{BKT})=2 T^\alpha_\mathrm{BKT}/\pi$ and $\alpha = 3$ are expected. The state with $\alpha = 1 \: (i.e.\: V \propto I)$ means normal state. Thereby, we can determine $T^\alpha_\mathrm{BKT}$ by measuring $\mathrm{V}(I)$ curve and estimating the value of $\alpha$ through fitting at various $T$.  \par

\begin{figure}[h]

      \includegraphics[width=\columnwidth]{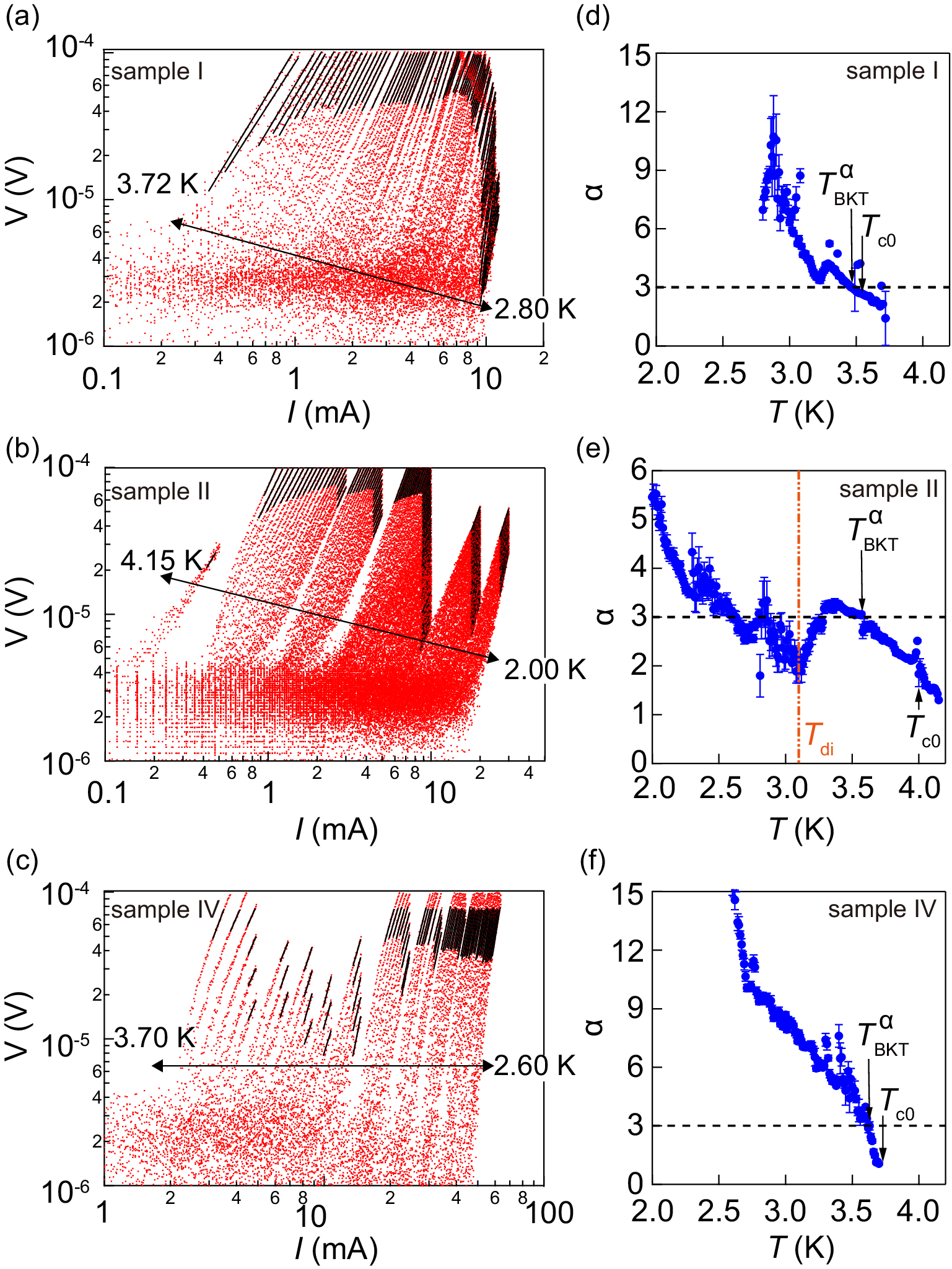}
      \caption{(a-c) Voltage-current ($\mathrm{V}(I)$) characteristics in samples I, II, and IV. The temperature spacing of each $\mathrm{V}(I)$ curve is 0.01\,K. The measurement range of each $\mathrm{V}(I)$ curve is determined to avoid Joule heating effects. The region of fitting $\mathrm{V}(I)$ curves to Eq. (\ref{eq:IV}) are shown as full lines in Figs. (a)-(c). (d-f) $\alpha(T)$ estimated by Eq. (\ref{eq:IV}) in samples I, II, and IV. In (e), the dash-dotted line indicates $T_\mathrm{di}=$ 3.10\,K, which is defined in Fig. \ref{fig:MT}(b). The dashed lines in Figs. (d)-(f) mean $\alpha = 3$, where the BKT transition occurs.}
      \label{fig:IV}

    \end{figure}

    Figures \ref{fig:IV}(a-c) show the temperature dependence of $\mathrm{V}(I)$ curves of samples I, II, and IV, respectively. The interval of the temperature is 0.01\,K. The non-linear behavior of the $\mathrm{V}(I)$ curves states that the zero resistivity state is attributed not to superconductivity but to the BKT state. We fitted $\mathrm{V}(I)$ curves using Eq. (\ref{eq:IV}) to determine $\alpha$. The fittings are carried out the region where $I$ is larger than the inflection point of $\mathrm{V}(I)$. This is to avoid the $I$ region where super- and normal currents mix. Temperature dependence of $\alpha$ is shown in Fig. \ref{fig:IV}(d-f) of samples I, II, and IV, respectively. $\alpha(T)$ decreases with temperature for samples I, II, and IV except near the inflection point. This reflects the decrease in the superfluid density \cite{Halperin1979,Venditti2019, Toyama2022}. The reason why $\alpha(T)$ is not monotonic in sample I and II may be attributed to the phase transition by the remaining superconducting volume. In Fig. \ref{fig:IV}(e), the dash-dotted line indicates $T_\mathrm{di}$ of this sample (see also Fig. \ref{fig:MT}(b)). Below $T_\mathrm{di}$, the diamagnetic signal dominates the magnetic response, and the coherency between superelectrons develops below $T_\mathrm{di}$. The dip point of $\alpha(T)$ coincides with $T_\mathrm{di}$. In $T_\mathrm{di}<T<T^\alpha_\mathrm{BKT}$, $J_s$ may have non-monotonic change against the temperature. For sample I, although $T_\mathrm{di}$ cannot be determined by the magnetization measurement (Fig. \ref{fig:MT}(a)), the non-monotonic change is likely due to the same reason in sample II. When $T=$ 3.46\,K, 3.52\,K, and 3.63\,K, the BKT transition occurs with $\alpha = 3.01\pm0.02$, $3.04\pm0.04$, and $3.00\pm0.20$ in samples I, II, and IV, respectively. This temperature is within the expected range of $T^\mathrm{HN}_{\rm{BKT}}$ determined earlier from $\mathrm{\rho}(T)$ (see Fig. \ref{fig:RT_HN}(d-f) and Table \ref{table:BKT}). \par

    \begin{figure*}[!ht]

      \includegraphics[width=\textwidth]{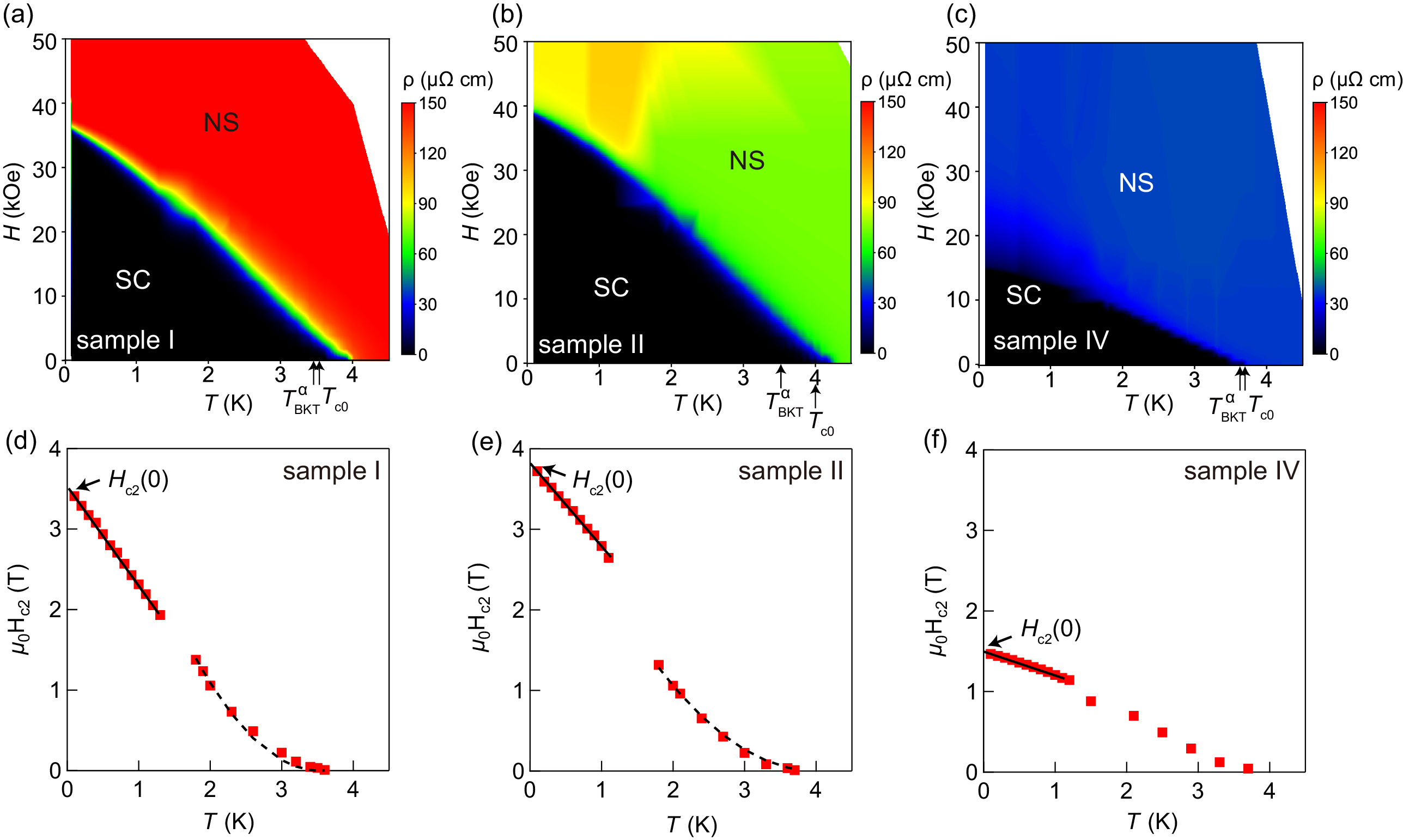}
      \caption{(a-c) Maps of the resistivity of samples I, II, and IV measured as a function of temperature and external magnetic field in the vicinity of superconductivity. The magnetic field is applied perpendicular to the sample surface. In the black region, $\mathrm{\rho}(T,H)=0$. NS represents the normal state. The color bar indicates the $\mathrm{\rho}(T,H)$ variation near the BKT transition. (d-f) Temperature dependence of $H_\mathrm{c2}$ in samples I, II, and IV. In (d)-(f), solid lines are linear fits. The dashed lines in (d) and (e) are fits to ($H_\mathrm{c2} \propto (1-T/T_\mathrm{c0})^2$).}
      \label{fig:RTH}

    \end{figure*}

    \subsection{Magnetoresistance measurement}
    The BKT transition exist only in the absence of an external magnetic field. If an external magnetic field is applied, the shift in resistivity should be considered as an effective reduction in the vortex-antivortex strength. We investigated $H$ and $T$ dependence of $\mathrm{\rho}(T,H)$ to examine the temperature dependence of the upper critical field $H_\mathrm{c2}$. Strictly speaking, the BKT phase does not have "upper critical field". Nevertheless, $H_\mathrm{c2}-T$  shows one of the phenomenological proofs of the BKT phase, where $H_\mathrm{c2}$ is defined by $\mathrm{\rho}(T, H_\mathrm{c2})=0$. Thus, we use $H_\mathrm{c2}$ to compare our results with the previous studies \cite{Toyama2022,Yamada2013,Matetskiy2015}. Figures \ref{fig:RTH}(a-c) show the $\mathrm{\rho}(T,H)$ mappings. The magnetic field is applied perpendicular to the film surface. The clear transition from the superconducting state (black region) to normal state is observed. This transition corresponds to $H_\mathrm{c2}$, and its temperature dependence from the $\mathrm{\rho}-H$ curves is shown in Fig. \ref{fig:RTH}(d-f) of samples I, II, and IV, respectively. $H_\mathrm{c2}$ decreases monotonically with temperature, reflecting the reduction in the vortex-antivortex strength. Below 1.1 K, $H_\mathrm{c2}$ is linear against the temperature. This is a typical trend for the BKT phase \cite{Toyama2022,Yamada2013,Matetskiy2015}. On the other hand, near $T\sim T_\mathrm{c0}$ in samples I and II, quadratic temperature dependence with positive curvature is dominant, and the fitting results by $H_\mathrm{c2} \propto (1-T/T_\mathrm{c0})^2$ are shown in the dashed lines. This quadratic behavior is also seen in sample III (see \cite{Sup}). BKT and Ginzburg-Landau (GL) theories do not give comprehensive representation that $H_\mathrm{c2}$ is linear in $T\sim 0$ and quadratic in $T\sim T_\mathrm{c0}$ \cite{Werthamer1966, Maki1964, Gennes1964, Maekawa1983, Smith2000}. On the other hand, the papers for multiband superconductivity argue that the power law dependence of $H_\mathrm{c2}$ in $T\sim T_\mathrm{c0}$ is attributed to defects and inhomogeneity \cite{Muller2001,Edge2015}. In our ReN$_x$ thin films, the two different superconducting phases of ReN and ReN$_2$ can coexist, and the model for the temperature dependence of $H_\mathrm{c2}$ in multiband superconductivity is applicable. Additionally, it is consistent with the fact that the quadratic temperature dependence is not observed in sample IV, which has relatively higher crystalline quality.\par
    The intercept of $H_\mathrm{c2}$ axis ($=H_\mathrm{c2}(0)$) by the linear fitting between 0.1 K and 1.1 K is estimated to obtain $H_\mathrm{c2}(0)$, as shown in Table \ref{table:Tc}. According to GL theory, we can discuss inhomogeneity in terms of the coherence length \cite{Yamada2013}.  Using $H_\mathrm{c2}(0)$, $\xi_\mathrm{GL}(0) = \sqrt{\Phi_0/2\pi H_\mathrm{c2}(0)}$, where $\Phi_0$ is flux quantum, can be estimated. In clean limit, $\xi_\mathrm{GL}(0)\sim\xi_0$ well below $T_\mathrm{c0}$. However, one can see $\xi_\mathrm{GL}(0)\ll\xi_0$ in Table \ref{table:Tc}, implying that the coherence length is dominated by $\ell$, and samples I, II, and IV are in dirty limit. Such condition of the samples helps distinguish $T_\mathrm{c0}$ with $T_\mathrm{BKT}$. Nevertheless, excessive inhomogeneity masks the BKT physics, and it can also cause the non-linear $\mathrm{V}(I)$ characteristics \cite{Venditti2019}. Inhomogeneous superconductors have islands with different strength of local superconducting condensates. The finite $I$ can turn the weak superconducting islands into normal ones, resulting in the non-linear $\mathrm{V}(I)$ curves at $T_\mathrm{c0}$ ($\alpha(T_\mathrm{c0}) > 1$). We take an interface of LaTiO$_3$/SrTiO$_3$ (LTO/STO) as an example to compare with our case \cite{Venditti2019}. The interface of LTO/STO shows superconductivity, and this conduction is regarded as 2D. Although the interface of LTO/STO is one of the candidates for the BKT transition, the experimental data of the $\mathrm{V}(I)$ measurement shows $\alpha(T_\mathrm{c0}) = 4.5$. This indicates that the BKT analysis by $\mathrm{V}(I)$ characteristics is not justified in the LTO/STO because of the excessive inhomogeneity. Notably, the NbN$_x$ thin film  shows $\alpha(T_\mathrm{c0}) = 2.7$. In our case, $\alpha(T_\mathrm{c0})=$ 2.69 $\pm$ 0.03, 1.83 $\pm$ 0.25 and 1.05 $\pm$ 0.05 in samples I, II, and IV, respectively. $\alpha(T_\mathrm{c0})$ decreases with $\ell$ and $\xi_\mathrm{GL}(0)$ (see Table \ref{table:Tc}). From the ReN$_x$ samples presented in this investigation, sample IV with $\alpha(T_\mathrm{c0})$ close to unity highlights advantages of the ReN$_x$ system.

    \section{Conclusion}
    We investigated the ReN$_x$ system as a possible materials system to study the BKT phase transition in superconducting systems. Magnetization measurements show a transition from fully diamagnetic thin films towards ReN$_x$ films with a superelectron response. Our analysis of electronic transport parameters concluded that the mean free path length can be effectively limited into a regime where $\ell\ll\xi_0$. This approach enabled us to determine critical parameters of the BKT transition in ReN$_x$. We confirm the hallmarks of the BKT transition in three samples with different superconducting volume by resistivity, $\mathrm{V}(I)$, and magnetization measurements. We employed the BMO theory to determine the $T_\mathrm{BKT}$ temperatures of ReN$_x$ subject to their synthesis parameters. To obtain more precise information about the transition of superfluid stiffness $J_s$ and the mean field critical temperature $T_\mathrm{c0}$, the penetration depth measurements and mutual inductance measurements methods are required \cite{Mondal2011, Mondal2012,Yong2013}. Nevertheless, the experimental data obtained by the magnetometry and the transport measurements prove that ReN$_x$ is a promising material to investigate BKT physics. Compared with the archetypal system of NbN$_x$ thin films, ReN$_x$ thin films do not need any additional treatment to introduce disorder and inhomogeneity. Furthermore, our ReN$_x$ shows nearly perfect linear relation of $\mathrm{V}(I)$ near $T_\mathrm{c0}$ ($\alpha(T_\mathrm{c0})=1.05$) while $\alpha(T_\mathrm{c0})=2.7$ in NbN$_x$ thin films \cite{Venditti2019}. Our result suggests that ReN$_x$ thin films can be tuned in a way that allows for sufficient inhomogeneity to instill a topological quantum phase transition like BKT rather than NbN$_x$. We conclude that ReN$_x$ is a novel platform for exploring the BKT physics and this work paves the way for potential applications of ReN$_x$ thin films in superconducting devices.

    \section*{Acknowledgment}
    The authors thank Takayuki Ikeda for his support in HAADF-STEM, TED and EDX measurements. The authors thank Hiroshi Irie for his inspiring discussion.

    \section*{References}
    \renewcommand{\bibsection}{}

    \bibliography{PrReN}

% The \nocite command causes all entries in a bibliography to be printed out
% whether or not they are actually referenced in the text. This is appropriate
% for the sample file to show the different styles of references, but authors
% most likely will not want to use it.
%\nocite{*}

\end{document}